\documentclass[twocolumn,aps,prl,amsmath,amssymb,superscriptaddress]{revtex4-1}
\usepackage{amsmath,oplotsymbl}
\usepackage{amssymb}
\usepackage{wasysym}
\usepackage{pifont}
\usepackage{graphicx,overpic} 
\usepackage{mathrsfs}

\newcommand*{\ellp}{\ell_{\rm P}}

\newcommand{\ve}[1]{\ensuremath{\mbox{\boldmath$#1$}}}

\def\smallint{\begingroup\textstyle \int\endgroup}

\usepackage{xcolor}
\newcommand{\obs}[1]{{#1}}
\newcommand*{\localTime}{{\mathscr{L}}}
\newcommand*{\pen}{{\varepsilon}}

\begin{document}

\title{One-parameter scaling theory for DNA extension in a nanochannel}
\author{E. Werner}
\affiliation{Department of Physics, University of Gothenburg, SE-41296 Gothenburg, Sweden}
\author{G. K. Cheong}
\affiliation{Department of Chemical Engineering and Materials
Science, University of Minnesota -- Twin Cities, 421 Washington Ave SE, Minneapolis, MN 55455, USA}
\author{D. Gupta}
\affiliation{Department of Chemical Engineering and Materials
Science, University of Minnesota -- Twin Cities, 421 Washington Ave SE, Minneapolis, MN 55455, USA}
\author{K. D. Dorfman}
\email{dorfman@umn.edu}
\affiliation{Department of Chemical Engineering and Materials
Science, University of Minnesota -- Twin Cities, 421 Washington Ave SE, Minneapolis, MN 55455, USA}
\author{B. Mehlig}
\email{bernhard.mehlig@physics.gu.se}
\affiliation{Department of Physics, University of Gothenburg, SE-41296 Gothenburg, Sweden}
\date{\today}

\begin{abstract}
Experiments measuring DNA extension in nanochannels are at odds with even the most basic predictions of current scaling arguments for the conformations of confined semiflexible polymers such as DNA. We show that a theory based on a weakly self-avoiding, one-dimensional \lq telegraph\rq{} process collapses experimental data and simulation results onto a single master curve throughout the experimentally relevant region of parameter space and explains the mechanisms at play.

\end{abstract}

\maketitle

As the carrier of genetic information, DNA plays a key role in biology. At the same time, recent advancements in fluorescence microscopy allow DNA to serve as a model polymer for investigating fundamental questions in polymer physics \cite{Shaqfeh2005,Latinwo2011}. Nowhere is this dual importance more apparent than in the problem of DNA confinement in a nanochannel \cite{Reisner2012,Dorfman2013,Dai2015}. When the radius of gyration of the DNA molecule is larger
than the channel width, it extends along the channel. This stretching lies at the heart of genome mapping in nanochannels \cite{Lam2012}. Here the stretched DNA molecules, usually greater than 150 kilobasepairs in length, contain fluorescent markers that reveal sequence-specific information with kilobasepair resolution. This new method serves as a complement to next-generation {\it de novo} DNA sequencing \cite{Jo2007,Lam2012,Michaeli2012}.  

Polymer confinement has been investigated for four 
decades, starting with the scaling arguments
of Daoud and de Gennes \cite{Daoud1977}. Yet there is to date no microscopic theory explaining the experimental data of recent genome-mapping experiments \cite{Jo2007,Lam2012,Michaeli2012,Reinhart2015} in narrow nanochannels. 
The difficulty is that the channels are too wide
to apply scaling arguments derived for strong confinement \cite{Odijk2008,Muralidhar2014}, yet too narrow for the scaling arguments and theory 
\cite{Odijk2008,Wang2011,Dai2014,Werner2014,Werner2015} in
wider channels to hold.

\begin{figure}[t]
\begin{center}
\begin{overpic}[clip,width=0.9\columnwidth]{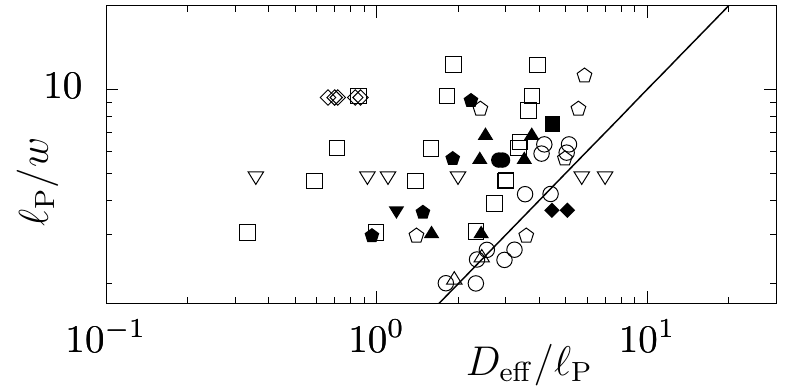}
\end{overpic}
\end{center}
\caption{Parameters for experiments on nanoconfined DNA: 
$\triangledown$ \cite{Reisner2005},
$\square$ \cite{Reisner2007},
$\blacksquare$ \cite{Thamdrup2008},
$\ocircle$ \cite{Zhang2008},
$\CIRCLE$ \cite{Utko2011},
$\triangle$ \cite{Kim2011},
$\blacktriangle$ \cite{Werner2012},
$\blacktriangledown$ \cite{Gupta2014},
$\Diamond$ \cite{Reinhart2015},
\ding{117} \cite{Gupta2015},
$\pentagon$ \cite{Iarko2015}, 
and $\pentagofill$ \cite{Alizadehheidari2015}.
For experiments using funnels \cite{Utko2011,Werner2012,Gupta2015} only maximum and minimum channel widths are indicated. The methods for selecting the 
data sets, and for computing the \lq effective channel width\rq{} $D_{\mathrm{eff}}$, $\ellp$, and $w$ from the experimental parameters, are described in the Supplemental Material~\cite{SM}. Solid line shows $D_{\rm eff}=\ellp^2/w$. }
\label{fig:pd}
\end{figure}

The challenge in developing a theory for the extension of nanoconfined DNA arises from its semiflexible nature. Semiflexible polymers are characterized by three length scales: the contour length $L$, the persistence length $\ellp$ quantifying the stiffness of the chain,
and the effective width $w$ that appears in the Onsager excluded volume \cite{Onsager1949}. For polyelectrolytes such as DNA, both the persistence length \cite{Odijk1977,Skolnick1977,Dobrynin2005,Trizac2016} and the effective polymer width \cite{Stigter1977} depend on electrostatic interactions.  Recent experiments are often conducted in high ionic-strength buffers. In this case $\ellp$ is approximately $50\,$nm \cite{Bustamante1994} while $w$ is around $5\,$nm \cite{Hsieh2008}, and thus $w \ll \ellp$.  This inequality emphasizes the intrinsic difficulty of describing DNA in a wide range of situations.  DNA is considerably stiffer than typical synthetic polymers, yet the number of persistence lengths $L/\ellp$ 
in genomic DNA samples is large.  Any theory for the conformational statistics of channel-confined DNA must account for  both the local stiffness of the polymer and excluded-volume interactions. This is a formidable challenge. Matters are further complicated by the fact that most recent genome-mapping experiments are performed in nanochannels that are about $50\,$nm wide \cite{Jo2007,Lam2012,Michaeli2012,Reinhart2015}. Essentially all experiments involving DNA (Fig.~\ref{fig:pd}) take place in channel sizes $D$ of the order of $\ellp$ and do not satisfy the criterion $D \gg \ellp^2/w$ required for the
scaling arguments of Ref.~\cite{Daoud1977} to  apply.

There is no microscopic theory for the extension of confined DNA for $D < \ellp^2/w$, despite numerous attempts \cite{Zhang2008,Cifra2009,Tree2013a}. Scaling arguments \cite{Wang2011,Dai2014,Werner2015} following Refs.~\cite{Daoud1977,Odijk2008} yield  the most useful description. If $D \gg \ellp$
they suggest that the extension $X$ scales as $X \sim D^{-2/3}$. The problem is that the 
theory fails when $D\approx \ellp$, and as a result it proves to be a poor description of most recent DNA experiments in nanochannels. 
The earliest experiments \cite{Reisner2005}, for instance, reported
a much larger exponent $X \sim D^{-0.85}$, and subsequent studies \cite{Persson2009,Gupta2014,Gupta2015} continue to report exponents exceeding the theoretical prediction.

\begin{figure}[t]
\begin{overpic}[clip,width=6cm]{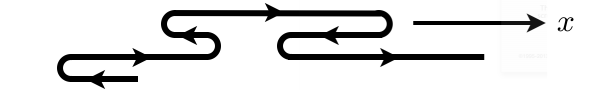}
\end{overpic}
\caption{Illustration of the telegraph process along the channel axis ($x$-axis). The walk is one-dimensional, but for clarity it is expanded vertically, to show the changes in direction that create  hairpin configurations of the confined DNA molecule.}
\label{fig:telegraph}
\end{figure}

We take a different approach in this Letter. We show that the DNA-confinement problem for $w\ll \ellp$ and $D\ll \ellp^2/w$ maps to the simple one-dimensional telegraph process in Fig.~\ref{fig:telegraph}, describing the correlated walk of a particle moving with velocity $v_0$ along the channel axis. The velocity changes sign at rate $r$, creating hairpin configurations in the particle path.  The process lasts from $t=0$ to $t=T$.  When the particle revisits a position it has previously visited, it incurs a penalty $\pen$.  
We show that this model 
collapses experimental and simulation data for the extension
throughout the experimentally relevant parameter range onto a {\em universal}  master curve as a function of a new scaling variable, 
$\alpha$, that measures the combined effects of confinement, local stiffness, and self avoidance.

We start by considering narrow channels, $D\ll \ellp$, and later extend the arguments to channel widths up to $\ellp^2/w$.  Since we want to compute the extension  of the DNA molecule along the channel axis, it suffices to consider the projection of the three-dimensional DNA configurations $\ve x(s)$ to the channel axis $x$. Here $s$ is the contour-length coordinate, 
it corresponds to time $t$ in the telegraph process. 
We decompose the probability $P[x(s)]$ of observing the projected conformation $x(s)$ as
\begin{equation}
\label{eq:p=p*A}
P[x(s)] \propto P_{\rm ideal}[x(s)]\, \mathscr{A}[x(s)]\,.
\end{equation}
The functional $P_{\rm ideal}[x(s)]$ is the probability of observing the conformation $x(s)$ in an ensemble of ideal confined polymers, disregarding self avoidance.  The functional $\mathscr{A}[x(s)]$ captures the effect of self avoidance. It equals the fraction of three-dimensional polymer configurations corresponding to $x(s)$ that contain no segments that collide with any other polymer segment. 

\begin{figure}[t]
 \begin{overpic}[clip,width=\columnwidth]{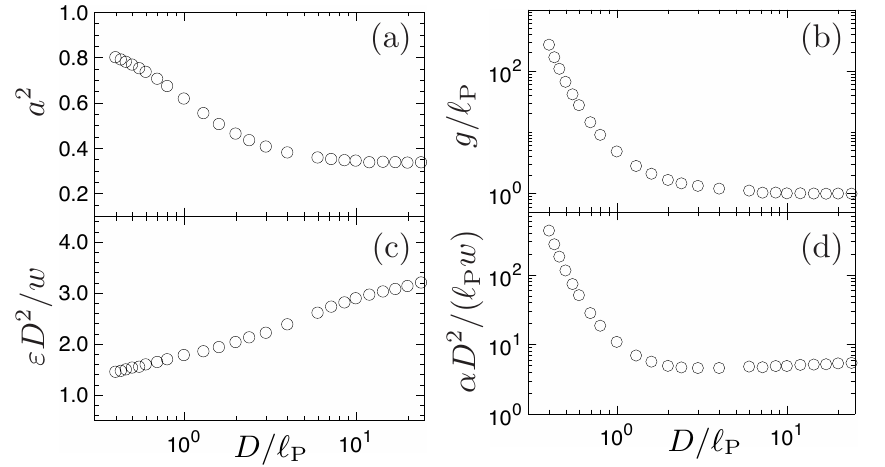}
 \end{overpic}
 \caption{
 \label{fig:alpha}
  Results from ideal-polymer simulations \cite{SM} showing how
$a=v_0$, $g = 1/(2r)$, $\pen$, and $\alpha$ depend on $D/\ellp$.}
 \end{figure}

When $D < \ellp$, the local conformation statistics are determined by Odijk's theory for narrow channels \cite{Odijk1983}, while the global statistics are dominated by a random sequence of direction changes (hairpins) \cite{Odijk2006}. Neglecting self-avoidance, the separation between neighboring hairpin-bends is 
exponentially distributed \cite{Odijk2006}. On length scales much larger than the deflection length \cite{Odijk1983} $\lambda\equiv (\ellp D^2)^{1/3}$ the central-limit theorem assures that local alignment fluctuations are negligible \cite{Odijk2008}. These two facts imply that the ideal problem maps onto the one-dimensional telegraph process in Fig.~\ref{fig:telegraph}. 
The correlation function of $v_x(s)$, the
channel-axis component of the tangent vector of the ideal polymer
decays exponentially \cite{SM}
\begin{equation}
\label{eq:cs}
\langle v_x(s) v_x(0) \rangle = a^2 \exp(-s/g)\,.
\end{equation}
The telegraph velocity has similar correlations:
\begin{equation}
\label{eq:corr}
\mbox{}\!\!\!\!\!\!\!
\langle v(t) v(0)\rangle = v_0^2\exp(-2rt)\,.
\end{equation}
Comparing Eqs.~(\ref{eq:cs}) and (\ref{eq:corr}) we see that the contour parameter $s$ maps to the time $t$ in the telegraph process, whereupon the polymer-contour length $L$ maps to the total time $T$ in the telegraph model.
The parameter $a$ quantifies the tendency of the tangent vectors to align with the channel \cite{Werner2012}. 
The parameter  $g$ is the global persistence length \cite{Odijk2006}, characterizing the typical distance between hairpin turns. 
These parameters map to those of the telegraph process as follows:
$a=v_0$ and  $g = (2r)^{-1}$. 
We measured how $a$ and $g$ depend on the physical parameters of the full
three-dimensional problem from simulations of confined ideal polymers. It turns out
that it suffices to determine just two curves (Fig.~\ref{fig:alpha}{\bf a} and {\bf b}), since $a$ and $g/\ellp$ depend on $D/\ellp$ only. 

Now consider the effect of self avoidance. In general it is very difficult to derive an expression for $\mathscr{A}[x(s)]$. But
for a weakly self-avoiding polymer, the problem simplifies considerably when the channel 
is so narrow that interactions between the polymer and the channel wall cause the three-dimensional configurations to lose correlations. 
We show in the Supplemental Material~\cite{SM} that 
\begin{equation}
\label{eq:ns}
\mathscr{A}[x(s)] \propto {\rm exp}\big[-\tfrac{\pen}{2}\smallint\!{\rm d}x\, \localTime^2(x) \big] \,.
\end{equation}
if $w\ll \ellp$.
Here $\localTime(x) \, {\rm d}x$ is the total amount of contour in the interval $[x,x + {\rm d}x]$ \cite{vanderhofstad2003}. 
The parameter $\pen$ penalizes overlaps. It is determined by the probability that two short polymer segments overlapping in  one dimension collide in  three dimensions \cite{SM}:
\begin{equation}
\label{eq:penalty} 
\pen = \langle \delta(y - y')\delta(z - z') v_{\rm ex}\rangle/\ell^2\,.
\end{equation}
The average is
over the conformations of the confined {\em ideal} polymer, and $y$ and $z$ are the transverse channel coordinates of a short polymer segment of length $\ell$.  Primed coordinates belong to a second, independent segment,
and $v_{\rm ex}$ is the excluded volume. 
The excluded volume depends on the segment orientation. If $\ell \gg w$, 
we have $v_{\rm ex} = 2w\ell^2\sin\theta$, where $\theta$ is the angle between the two segments  \cite{Onsager1949}. 
 Fig.~\ref{fig:alpha}{\bf c} shows how $\pen$ depends on $D/\ellp$, obtained by evaluating the average in 
(\ref{eq:penalty}) from three-dimensional simulations of confined ideal polymers \cite{SM}. A single curve is sufficient
to determine how $\pen$ depends on the physical parameters,
because $\pen D^2/w$ is a function of $D/\ellp$ only
(Supplemental Material \cite{SM}).

In the telegraph model self-avoidance is incorporated in the same way. 
Here, $\mathscr{L}$ has units of time/position. Eq.~(\ref{eq:ns}) then shows that $\pen$ has units of position/time$^2$. Since $r$ has units of time$^{-1}$, and $v_0$ of position/time, the only dimensionless combination of $\pen$, $r$, and $v_0$ is
\begin{equation}
\label{eq:alpha}
\alpha \equiv \pen/(2v_0 r) \,.
\end{equation}
In the limit of large $T$ only $\alpha$ can have physical significance.
Invoking our mapping between telegraph model and polymer problem we conclude that $\alpha$ is given by
\begin{equation}
\label{eq:alpha_dna}
\alpha = \pen g/a\,.
\end{equation}
This parameter measures the expected number of overlaps between the two strands of a hairpin of length $g$.

Eq.~(\ref{eq:alpha_dna}) has two important consequences. 
First, Eq.~(\ref{eq:alpha_dna}) allows us to generalize the mapping to all channel widths up to $\ellp^2/w$. To show this, consider first the ideal part.  Strictly speaking, the simple picture outlined above breaks down when $D \sim\ellp$ because the typical hairpin length $g$ becomes of the same order as $\ellp$.
But consider how $\alpha$ changes as $D$ approaches $\ellp$ from below. For
$w\ll\ellp$, the parameter $\alpha$ decreases below unity before $g=\ellp$ is reached,
and for small $\alpha$ the precise nature of the local conformations is irrelevant.
All that matters is that the ideal part is a diffusion process with exponentially decaying correlations of $v_x(t)$.  Similarly, the local probability of collision is still $\tfrac{\pen}{2}\,\localTime^2(x) {\rm d}x$, because each segment pair collides independently. The latter assumption eventually breaks down at $D \approx \ellp^2/w$ since the transversal 
segment coordinates become correlated. But up to this point
Eq.~(\ref{eq:ns}) is valid, as is Eq.~(\ref{eq:penalty}). 

Second, observables that are dimensionless in the telegraph model can only depend on $\alpha$, Eq.~(\ref{eq:alpha_dna}), in the limit of large $L$.
This combination, $\alpha$, is
plotted in Fig.~\ref{fig:alpha}{\bf d}. It turns
out that $\alpha D^2/(\ellp w)$ depends only on $D/\ellp$ \cite{SM}.
Now consider the average  extension, $X$, and the variance about that average, 
$\sigma^2$. In the telegraph model these quantities have units of  position and position$^2$, 
and for large values of $L$ they must be proportional to $L$. We therefore
conclude that the data must collapse as 
\begin{equation}
\label{eq:universalScalingFunctions}
X/(La) =f_X(\alpha)\quad\mbox{and}\quad \sigma^2/(Lga^2)=f_\sigma(\alpha)\,.
\end{equation}
Here $f_X$ and $f_\sigma$ are {\em universal} scaling functions  that depend only on $\alpha$.  
 \begin{figure*}[t!]
 \begin{overpic}[width=6.5in]{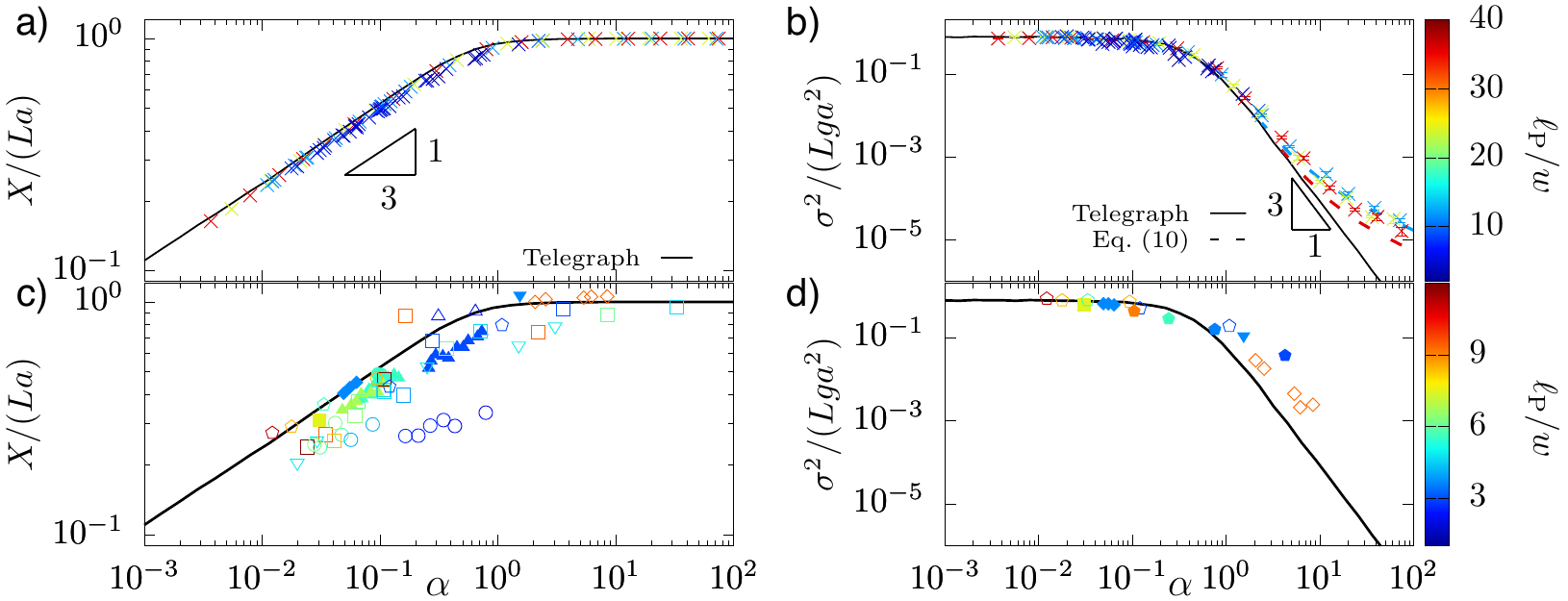}
 \end{overpic}
\caption{
 \label{fig:results}
One-parameter scaling of the mean extension $X$ and the extension variance $\sigma^2$. 
Comparison of one-parameter theory (solid black lines) to results of three-dimensional direct numerical simulations (DNS) (a,b) and experiments (c,d).
DNS: crosses. The DNS method \cite{Tree2013,Tree2013a} is described in the Supplemental Material~\cite{SM}.
Experiments (same as in Fig.~\ref{fig:pd}): 
$\triangledown$ \cite{Reisner2005},
$\square$ \cite{Reisner2007}, 
$\blacksquare$ \cite{Thamdrup2008}, 
$\ocircle$ \cite{Zhang2008} , 
$\CIRCLE$ \cite{Utko2011},  
$\triangle$ \cite{Kim2011},  
$\blacktriangle$ \cite{Werner2012},  
$\blacktriangledown$ \cite{Gupta2014},  
$\Diamond$ \cite{Reinhart2015},  
\ding{117} \cite{Gupta2015},  
$\pentagon$ \cite{Iarko2015},   
and $\pentagofill$ \cite{Alizadehheidari2015}. 
The details of these experiments and the selection of experimental data sets are described in the Supplemental Material~\cite{SM}.  
In addition, the predicted scalings for the mean extension, $X \sim \alpha^{1/3}$,
and for the extension variance, $\sigma^2 \sim \alpha^{-3}$, are indicated. The color bar shows
the range of $\ellp/w$ for DNS (top) and experiments (bottom). The dashed lines in panel (b) show theoretical predictions from Eq.~(\ref{eq:correction}) for $\ellp/w=12$ (dashed blue line) and $36$ (dashed red line). See Fig.~S-2 \cite{SM} for the telegraph-model results as a function of channel width $D$.}
\end{figure*}
We can numerically compute the form of these functions simply by simulating the telegraph model~\cite{SM}.

We have compared our theory to direct numerical simulations (DNS) of three-dimensional, confined, self-avoiding  wormlike chains  \cite{Tree2013a} using the PERM algorithm \cite{Grassberger1997,Prellberg2004}. 
Figs.~\ref{fig:results}{\bf a},{\bf b} show that our theory quantitatively captures the 
DNS results for all values of $\ellp/w$ tested \cite{SM}, up to $w/\ellp = 0.4$. 
This agreement is remarkable, as the theory assumes weak self avoidance, $w\ll \ellp$.

Figs.~\ref{fig:results}{\bf c} and {\bf d} show the comparison between the results of the experiments summarized in Fig.~\ref{fig:pd} and our theory. The theory not only collapses the experimental data, but provides good quantitative agreement, in particular with the most recent experiments \cite{Gupta2015,Reinhart2015,Iarko2015}. There is some scatter of the experimental data around the theoretical curve, but this is expected because the nanofluidic experiments are quite difficult to control.  

In the limit $\alpha\ll 1$ our theory allows to map the problem to 
an uncorrelated  weakly self-avoiding diffusion process \cite{van_der_hofstad_constants_1998,vanderhofstad2003}. This follows from the fact that the correlation function in the telegraph model, Eq.~(\ref{eq:corr}), decays to zero before the next collision occurs, for 
$\alpha\ll 1$. As a result the ideal random walk is simply diffusive, with diffusion constant $\mathscr{D} = v_0^2/(2r)$. 
This has two consequences. 

First, for $\alpha\ll 1$ observables depend on $v_0$ and $r$ only through the combination $\mathscr{D}$. Since the extension is linear in $L$ for large $L$, we deduce that $X/L$ can only depend on $\pen$ and $\mathscr{D}$ in this limit. Since $\mathscr{D}$ 
has units of position$^2$/time while $\pen$ has units of position/time$^2$ in the telegraph model, we see that the only possible
combination is $X/L\propto (\pen\mathscr{D})^{1/3}$. 
This gives $X/(La) = f_X(\alpha) \propto \alpha^{1/3}$, explaining the power law in Fig.~\ref{fig:results}{\bf a} and {\bf c}.
For the variance
we conclude that $\sigma^2/L \propto \mathscr{D}$, independent of $\pen$
(Fig.~\ref{fig:results}{\bf b} and {\bf d}). 
Alternatively we can deduce these scalings
by a mean-field argument, indicating that fluctuations of $\mathscr{L}$ 
are negligible when $\alpha \ll 1$. Assuming that $\mathscr{L} \sim T/X$ 
we find for the extension distribution $P(X)\sim \exp[-rX^2/(2v_0^2T)-\tfrac{\pen}{2}T^2/X]$, whereupon ${\rm d}\log P/{\rm d}X=0$ yields $X /(v_0 T) \sim \alpha^{1/3}$. For the variance we obtain $\sigma^2 r/(v_0^2 T) \sim \alpha^0$.

Second, we can use the exact mathematical
results derived in Ref.~\cite{van_der_hofstad_constants_1998}
to deduce the prefactors:
\begin{subequations}
\begin{align}
\label{eq:universalAsymptotes}
f_X(\alpha) &= c_X \alpha^{1/3} \quad (1.104 \le c_X \le 1.124)\,, \\
f_\sigma(\alpha) &= c_\sigma\,    \quad\quad\quad (0.72 \le c_\sigma \le 0.87)
\end{align}
\end{subequations}
as $\alpha\to 0$. The constraints for $c_X$ and $c_\sigma$ 
are rigorously proven mathematical bounds
\cite{van_der_hofstad_constants_1998}.

Now consider the limit of large $\alpha$. 
The extension $X$ tends to $La$ in this limit \cite{Odijk2008},
since the frequency of hairpins tends to zero. 
The variance decays as  $\sigma^2\sim \alpha^{-3}$, as Fig.~\ref{fig:results} shows. To deduce this power law, we estimate the variance of the strongly extended polymer as (number of hairpins) $\times$ (hairpin extension)$^2$.
To determine the number of hairpins, note that the expected number of collisions for a hairpin of contour length $h$ is $\sim \alpha\, h / g$. 
The resulting hairpin is therefore likely to survive the collision check only if $h$ is of the order $g /\alpha$ or smaller. This requires a second switch of direction within the length $g /\alpha$. 
This occurs with probability $(g / \alpha) / g = \alpha^{-1}$, so that the number of hairpins is $(L / g) \alpha^{-1}$. 
To obtain the hairpin extension we multiply its contour length $h\sim g/\alpha$ by its alignment $a$, so that the typical hairpin extension becomes $\sim g a /\alpha$.
Therefore $\sigma^2 \propto (L / g) \alpha^{-1}  \times (g a /\alpha)^2 = L g a^2 \alpha^{-3}$ for large $\alpha$.  

For very large values of $\alpha$, the theory fails [Fig.~\ref{fig:results}(b)] because hairpins are so rare that alignment fluctuations (not included in the telegraph model) dominate the variance \cite{Odijk2008}. 
This correction is taken into account simply by adding the
variance in the extreme Odijk limit \cite{Burkhardt2010},  
\begin{equation}
\label{eq:correction}
\sigma^2 =  d_\sigma (L ga^2) \alpha^{-3}+\sigma_{\rm Odijk}^2 \quad \mbox{as}\quad \alpha\to \infty\,.
\end{equation}
Here $d_{\sigma}$ is a universal constant. By fitting the solid line in Fig.~\ref{fig:results}(b) for $\alpha > 10$ we find $d_{\sigma} \obs{\approx 0.09}$. 
We observe excellent agreement between this refined theory and the simulation data for $\ellp/w$ = 12. For the stiffer polymers ($\ellp/w$ = 36) still longer contour lengths are required to reach the large-$L$ limit and to reduce the statistical error.

Finally we show that our theory
contains scaling laws derived earlier as particular asymptotic limits.
In very narrow channels, $a \approx 1$ 
and $\langle \sin \theta \rangle \approx (D/\ellp)^{1/3}$. Using these approximations in Eqs.~(\ref{eq:penalty}) and (\ref{eq:alpha}) gives 
\begin{equation}
\alpha =Cg w \,(D^5 \ellp)^{-1/3} = C\xi \quad\quad(D\ll \ellp)\,,
\label{eq:alpha_BFO}
\end{equation}
where $C\approx 1.95$ is a constant \cite{SM}.
The parameter  $\xi$ appears in Odijk's scaling theory \cite{Odijk2008} and the extension scales as $X \sim \xi^{1/3}$ \cite{Odijk2008} in this special limit. 
In wide channels, for $\ellp \ll D \ll \ellp^2 / w$, 
we have that $a = 1/\sqrt{3}$ and $g \approx \ellp$. Using diffusion approximations 
for the distribution of the polymer in the channel \cite{SM} gives
\begin{equation}
\alpha = 9\sqrt{3}\pi\,w \ellp/(8D^2) \ll 1 \quad (\ellp \ll D \ll \ellp^2 / w)\,.
\label{eq:alpha_edG}
\end{equation}
This is the result of Ref.~\cite{Werner2014}, implying the same scaling $X \sim D^{-2/3}$ that Odijk's scaling arguments \cite{Odijk2008} predicts in this asymptotic limit. 
At first glance it is perhaps surprising that the two distinct limits 
(\ref{eq:alpha_BFO}) and (\ref{eq:alpha_edG}) are described by the same random-walk process. After all, the three-dimensional polymer conformations are entirely different in the two regimes,
described by invoking deflection segments, hairpins, and blobs. Our universal theory, by contrast, rests on the fact that the macroscopic statistics of a weakly interacting 
random walk do not depend on the microscopic details of the process \cite{Grosberg1994}. 

We can also conclude that the DNA experiments shown in Fig.~\ref{fig:pd} cannot obey
the scalings $X\sim D^{-2/3}$ or $X \sim \xi^{1/3}$ because the experiments do not satisfy the 
strong inequalities $D\gg\ellp$ or $D\ll \ellp$ (see Fig.~S-5 in the Supplemental Material \cite{SM}),
and therefore do not reach the asymptotic limits required for these
power laws to emerge. Our theory shows, and Fig.~\ref{fig:results}(a) confirms that $X \sim \alpha^{1/3}$ for small values of $\alpha$. 
But the parameter $\alpha$ depends upon $D$ and $\ellp$ in an intricate way via Eq.~(\ref{eq:alpha_dna}), in general not in a power-law fashion.

In summary, we have shown that DNA confinement in nanochannels can be modeled by a telegraph process, collapsing all of the data in terms of a scaling variable $\alpha$.
Our theory brings to the fore universal properties of confined polymers in a good solvent in a way that is obscured by the prevailing scaling theories \cite{Daoud1977,Odijk1983,Odijk2008,Werner2014,Muralidhar2014,Dai2014}. The distinction between deflection segments, hairpins, and blobs, which leads to the need to define separate regimes is not necessary. Rather, the statistics of the confined polymer chain for $D \lesssim \ellp^2/w$ adopt a universal behavior at sufficiently long length scales, independent of the details of the microscopic model. 
\nocite{werner2013}

\begin{acknowledgements}
\obs{We thank Daniel \"Odman for helping us to uncover an error in the simulations 
of the telegraph model.}
This work was supported by VR grant 2013-3992 and by the National Institutes of Health (R01-HG006851). DG acknowledges the support of a Doctoral Dissertation Fellowship from the University of Minnesota.  Computational resources were provided by the Minnesota Supercomputing Institute, and by C3SE and SNIC.  
\end{acknowledgements}

%
\end{document}


\noindent \large{\bf Supplemental Material for ``One-parameter theory for DNA extension in a nanochannel''}

\noindent E. Werner,$^1$ G. K. Cheong,$^2$ D. Gupta,$^2$ K. D. Dorfman,$^{2,a}$ and B. Mehlig$^{1,b}$

\noindent $^1$ Department of Physics, University of Gothenburg, SE-41296 Gothenburg, Sweden

\noindent $^2$ Department of Chemical Engineering
and Materials Science, University of Minnesota -- Twin Cities,
421 Washington Avenue SE, Minneapolis, Minnesota 55455, USA

\noindent $^a$ Email: dorfman@umn.edu

\noindent $^b$ Email: bernhard.mehlig@physics.gu.se

\tableofcontents

\clearpage

\section{\,\,Telegraph process}       
Eqs.~(1) to (7) in the main text show how the problem of computing the extension of a DNA molecule
in a nanochannel is mapped to a one-dimensional telegraph process. In this Section we give all details necessary to derive this mapping,
to evaluate asymptotic limiting cases, and to simulate the process.

\begin{figure}[t]
\begin{center}
\begin{overpic}[width=8cm]{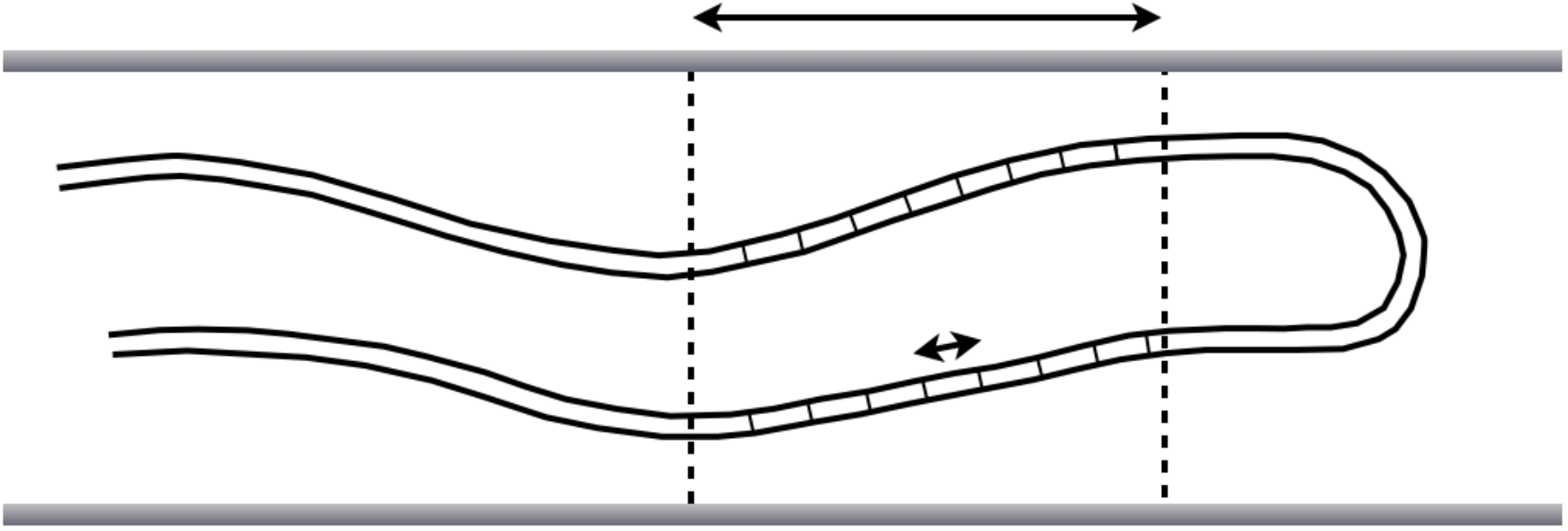}
\put(59,13){$\ell$}
\put(57,35){${\rm d}x$}
\end{overpic}
\end{center}
\caption{
\label{fig:twostrands} Illustration of two overlapping strands of 
a DNA molecule confined to a narrow channel. Discretization of the problem
to derive the telegraph model, $w \ll \ell \ll {\rm d}x$. Here $w$ is the effective width of the DNA molecule, $\ell$ is the length of a short DNA segment,
and ${\rm d}x$ is the discretization in the channel direction.} 
\end{figure}

\subsection{\,Derivation of Eq.~(4) in the main text}      
Consider a weakly self-avoiding polymer ($w \ll D$) in a narrow channel ($D \ll \ellp$).  We divide the  channel into slices of width ${\rm d}x$
and assume that ${\rm d}x \gg \lambda$, where $\lambda\equiv (\ellp D^2)^{1/3}$ is the Odijk deflection length,\cite{Odijk1983} so that we can neglect fluctuations in the contour length and alignment of a hairpin strand. We also assume that
${\rm d}x$ is small enough so that $p_{\rm coll}(x) {\rm d}x\ll 1$.
Here $p_{\rm coll}(x) {\rm d}x$ is the probability that one or more collisions occur in the channel segment between $x$ and $x+{\rm d}x$. This assumption assures
that we can neglect higher-order terms in a Taylor expansion below. 
Provided that  $w \ll D$ it is possible to satisfy both inequalities simultaneously for all configurations 
$x(t)$ that have non-negligible weight in the ensemble.
The probability $\mathscr{A}[x(t)]$ that the polymer configuration is free of overlaps is given by
\begin{align}
\mathscr{A}[x(t)] &= \hspace*{-3mm}\prod_{\mbox{\small all slices}}\hspace*{-3mm} 
\big(1 - p_{\rm coll}(x)\,{\rm d}x\big) 
= {\rm exp}\big\{\hspace*{-3mm}\sum_{\mbox{\small all slices}} \hspace*{-3mm}\log\big[1 - p_{\rm coll}(x){\rm d}x
\big] \big\}\,.
\nonumber
\end{align}
Since $p_{\rm coll}(x) {\rm d}x\ll 1$ this expression equals
\begin{align}
&= {\rm exp}\big\{\,-\hspace*{-3mm}\sum_{\mbox{\small all slices}} \hspace*{-3mm}p_{\rm coll}(x){\rm d}x\big\} 
 = {\rm exp}\big\{-\smallint{\rm d}x\,\, p_{\rm coll}(x)\big\}\,. 
\label{eq:Apcoll}
\end{align}
To evaluate the integral, we consider the two overlapping strands as shown in Fig.~\ref{fig:twostrands}. What is the probability that they collide in the interval between $x$ and $x+{\rm d}x$? To answer this question we divide each strand into many short
segments of length $\ell$. We must assume that $w \ll \ell$,
so that collision checks between neighboring segments are independent.  In the interval from $x$ to $x+{\rm d}x$ there  are 
 $(L_{\rm s} /\ell)^2$ segments pairs to check, $(L_{\rm s} /\ell)$
on each strand. Here $L_{\rm s}$ is the average strand length in the interval. 
Assume that the interval contains $N_{\rm s}$ strands.
Then the probability that there is at least one collision
in the interval from $x$ to $x+{\rm d}x$ is
\begin{equation}
\label{eq:pcoll}
p_{\rm coll} (x)\,{\rm d}x  =  p\,\, \Big(\frac{L_{\rm s}}{\ell}\Big)^2 N_{\rm s}(N_{\rm s}-1)/2 \,.
\end{equation}
Here $p$ is the probability that two randomly chosen short segments collide. Now we introduce the \lq local time\rq\cite{vanderhofstad2003} $\localTime(x)$, where $\localTime(x){\rm d}x$ is defined as the total contour length of the configuration $[x(t)]$ occupying $[x,x+{\rm d}x]$. 
It follows from this definition that $\localTime$ is in fact a density, it has units time/position. 
In terms of $\localTime(x)$ Eq.~(\ref{eq:pcoll}) becomes:
\begin{equation}
\label{eq:pcoll_tau}
p_{\rm coll} (x)\,{\rm d}x = \frac{p\diff x}{2\ell^2}\, \big[\localTime^2(x) \,{\rm d}x - 
\localTime(x) L_{\rm s} \big]
= \frac{\pen}{2}\, \big[\localTime^2(x) \,{\rm d}x -
\localTime(x) L_{\rm s} \big]\,,
\end{equation}
where we have introduced the \lq penalty\rq{} parameter 
\begin{equation}
\label{eq:pen_definition}
\pen \equiv \frac{p \,{\rm d}x}{\ell^2}\,.
\end{equation}
It has units of position/time$^2$.  Upon integration over $x$ 
the first term gives Eq.~(4) in the main text:
\begin{equation}
\mathscr{A}[x(t)] =\mathscr{A}_0\, {\rm exp}\big\{-\tfrac{\pen}{2}\smallint\!{\rm d}x \,\localTime^2(x)\big\}\,.
\label{eq:pcoll_tau_2}
\end{equation}
The second term gives the normalization factor $\mathscr{A}_0$. We evaluate it using $L_{\rm s} = \diff x/a$:
\begin{equation}
\label{eq:A0}
\mathscr{A}_0 = {\rm exp}\big\{\tfrac{\pen L}{2a}\big\}\,.
\end{equation}
Here $L$ is the contour length of the polymer.
This concludes the derivation of Eq.~(4) in the main text for $D\ll\ellp$. 
Our derivation assumes that different channel slices can be treated
independently, and that segment pairs within a slice are independent.
The first assumption is satisfied in narrow channels 
($D\ll \ellp$) since ${\rm d}x\gg \lambda$. The second assumption holds
since $p_{\rm coll}(x){\rm d}x \ll 1$ because this condition ensures that the result
of a collision check for a pair of segments is independent of the result
of checking any other segment pair. 

Now consider what happens when $D$ increases. Fig~3(b) in the main text shows that the global persistence length decreases as $D$ grows,
and the picture underlying the derivation outlined above breaks down when $g=\ellp$ because the notion of well aligned strands no longer applies 
in this case. This is, however, of little consequence because the universal scaling parameter 
$\alpha$ decreases below unity before $g=\ellp$ is reached.
For small values of $\alpha$, the correlations of the tangent vectors decay before
a collision occurs. For large-contour length separations the tangent vectors thus
perform an uncorrelated random walk with diffusion constant $\mathscr{D}=a^2g = v_0^2/(2r)$. Local conformations are no longer described by hairpins, but the precise nature of the local conformations is irrelevant in this limit, because
the macroscopic conformation statistics  of a self-avoiding polymer 
do not depend on the microscopic model.\cite{Grosberg1994} 
The ideal probability is simply Gaussian at contour-length scales above the
global persistence length $g$. 
The telegraph process can still be used because it gives precisely this ideal distribution in the limit of $\alpha \ll 1$.

Now consider the effect of self-avoidance.  
As $g\to \ellp$ the local conformations no longer resemble hairpins, so
that the number of hairpin strands ceases to be well defined. But 
Eq.~(\ref{eq:pcoll_tau_2}) remains valid, because
this equation simply expresses the fact that the probability of a collision
in a channel slice is proportional to the number of segment pairs in that slice.
Eventually, however, Eq.~(\ref{eq:pcoll_tau_2}) must break down,
when transversal segment coordinates become correlated. This
occurs for $D\gg \ellp^2/w$. 

\subsection{\,Derivation of Eq.~(5) in the main text}      
Recall that $p$ denotes the probability that two short segments of length $\ell$ collide, given that they occupy the same channel slice of width ${\rm d}x$. Denote the positions of the two segments by ${\ve x} = (x, y, z)^{\sf T}$ and ${\ve x}' = (x', y', z')^{\sf T}$, and the tangent
vectors  by ${\ve v}$ and ${\ve v}'$. The collision probability $p$ can be written as 
\begin{align}
p = \langle \chi(\ve x - \ve x', {\ve v}, {\ve v}')\rangle\,.
\end{align}
Here $\langle \ldots \rangle$ is an average over
$\ve x,{\ve v}$ and $\ve x',{\ve v}'$, distributed with the independent 
probabilities
$p(\ve x,{\ve v})$ and $p(\ve x',{\ve v}')$ of the two segments in the interior of the polymer. \cite{werner2013}
The indicator function $\chi(\ve x - \ve x', {\ve v}, {\ve v}')$
equals unity when the two segments collide, and zero otherwise. 
Since $D\gg \ell$, the probability density $p(\ve x,\ve v)$ varies little as $\ve x$ changes over the
length of the segment. We can therefore write
\begin{align}
\label{eq:chi_coll}
\chi = \delta(\ve x - \ve x') v_{\rm ex}({\ve v}, {\ve v}')\,,
\end{align}
where $v_{\rm ex}({\ve v}, {\ve v}')$ is the volume of the region surrounding a segment with tangent $\ve v$ that is excluded (on average) to other
segments that have tangent vector $\ve v'$.  For $\ell\gg w$ we can use Onsager's result \cite{Onsager1949} for the excluded volume:
\begin{align}
v_{\rm ex}({\ve v}, {\ve v}') = 2\ell^2 w \sin\theta(\ve v,\ve v')\,.
\end{align}
Here $\theta(\ve v,\ve v')$ is the angle between
the two segments with tangents $\ve v$ and $\ve v'$.
Finally, we note that the distributions of $x$ and $x'$ are simply uniform in the slice of width ${\rm d} x$. This yields 
\begin{equation}
p = \langle \delta(y-y') \delta(z-z') v_{\rm ex}({\ve v}, {\ve v}')\rangle / {\rm d}x,
\end{equation}
or, using the definition (\ref{eq:pen_definition}) of $\pen$, 
\begin{equation}
\label{eq:penresult}
\pen = \langle \delta(y-y') \delta(z-z') v_{\rm ex}({\ve v}, {\ve v}')\rangle /\ell^2\,.
\end{equation}
This is Eq.~(5) in the main text. The derivation
did not make any assumptions about the channel width, $D$. This means
that (\ref{eq:penresult}) is valid throughout the regime of validity
of (\ref{eq:pcoll_tau_2}).

\subsection{\,Asymptotic limits for large and small $\alpha$}
\label{sec:limits}
Eqs.~(11) and (12) in the main text describe asymptotic limits of the telegraph process.  
Here we give additional details for these two asymptotic limits. 

First when $\ell_p \ll D \ll \ellp^2/w$ then the correlation
function is the same as that of the unconfined DNA, wherein $a = 1/ \sqrt{3}$ and $g = \ellp$. \cite{Werner2014} 
To calculate $\pen$, we use a diffusion approximation\cite{werner2013}
for $p(\ve x_\perp,{\ve v})$, the joint distribution of $\ve x_\perp = (y,z)^{\sf T}$ and $\ve v$:
\begin{equation}
p(\ve x_\perp,{\ve v})=\frac{1}{4\pi} \frac{4}{D^2} \sin^2(\pi y / D) \sin^2(\pi z / D)\,.
\end{equation}
The factor of $(4\pi)^{-1}$ comes from the fact that angular orientations are isotropic
in this limit. Using these results in Eq.~(\ref{eq:penresult}) gives
\begin{equation}
\pen  = \frac{9\pi}{8} \frac{w}{D^2} \,.
\end{equation}
We can recover the results of Werner and Mehlig\cite{Werner2014} by setting the discretization scale $\ell$ to $2 \ellp$, whereupon
we obtain  Eq.~(3) of Werner and Mehlig:\cite{Werner2014}
\begin{equation}
\alpha = \frac{9 \sqrt{3} \pi}{8} \frac{w \ellp}{D^2}\,.
\end{equation}

Second when $D \ll \ellp$, the prefactor for the correlation function approaches $a = 1$. 
While there is no diffusion approximation for the Odijk regime, we know from dimensional analysis that the probability density of $\ve x_\perp$ must be of the form $p(\ve x_\perp) = D^{-2} h(y/D) h(z/D)$, and $\langle \sin(\theta)\rangle \approx (D/\ellp)^{1/3}$, see Ref.~\citenum{Odijk2008}. 
As a result we find from Eq.~(\ref{eq:penresult})
\begin{equation}
\label{eq:lexell}
\pen   \approx \frac{C w}{D^{5/3} \ellp^{1/3}}
\end{equation}
with an unknown numerical prefactor $C$. 
Fitting the factor $C$ in Eq.~(\ref{eq:lexell})
to the ideal-polymer simulations in Fig.~\ref{fig:parameters_ideal}(c) gives $C\approx 1.95$.  For  $\alpha$ we find the expression:
\begin{equation}
\alpha \approx C \frac{wg}{D^{5/3} \ellp^{1/3}}\,,
\label{eq:xi_Odijk}
\end{equation}
where the right-hand  side is proportional to the parameter $\xi$ defined by Odijk.\cite{Odijk2008} 
We emphasize that our telegraph theory does not contain any unknown factors. The numerical constant $C$
occurs in the relation between our exact theory and the parameter $\xi$ in the asymptotic limit $D\ll \ellp$.

\subsection{\,Computer simulations of telegraph model}
\label{sec:simulation_telegraph}
The solid lines in Fig.~4 in the main text were obtained by simulating a discretized version of the weakly self-avoiding telegraph model described in the main text, with $v_0=a=1$. We discretize the telegraph process into segments of length $\ell$ as 
follows.  If step number $j-1$ is in the positive (negative) direction, then step number $j$ is taken in the opposite direction with probability $p_{\rm switch} = r \ell$, and in the same direction with probability $1 - p_{\rm switch}$. The penalty 
is included by a probability 
$p_{\rm discard} = \pen \ell$ of discarding a configuration when a segment 
is in the same spot as a previous segment.
If there are $\tau$ segments at a given site, then the probability of surviving this check is $(1-p_{\rm discard})^{\tau} \approx \exp[-p_{\rm discard} \tau]$. 

For our simulations, we modified the algorithm described by Smithe {\it et al.}\cite{Smithe2015} to include correlations between steps. In this algorithm, a fixed number of chains grow in parallel. Each chain is grown by taking steps on a one-dimensional lattice. If step number $j-1$ is in the positive (negative) direction, then step number $j$ is taken in the same direction with probability $ 1 - p_{\rm switch}$, and in the opposite direction with probability $p_{\rm switch}$. If the step lands on a site that it has already visited $\tau$ times, the chain growth continues to step $j+1$ with probability $\exp(-p_{\rm discard} \tau)$, and is discarded with probability $1-\exp(-p_{\rm discard} \tau)$. If a chain is discarded, then one of the other chains is duplicated in order to keep the number of chains constant. Details of how to calculate weighted averages from this algorithm are described elsewhere.\cite{Dai2014,Smithe2015}
To obtain the master curve for the mean and variance of the extension 
shown in Fig.~4 in the main text we performed simulations of the telegraph model, using the algorithm described above. For these simulations we 
set $p_{\rm discard}$ to $0.1$, $0.01$, or $0.001$, and varied $p_{\rm switch}$ in logarithmic steps from $2^{-1}$ to $2^{-11}$. 

Now consider how to determine the scaling parameter $\alpha$
for the discretized process. Recalling that $v_0 = 1$, we have
\begin{equation}
\label{eq:ads}
 \alpha \equiv\frac{\pen}{2 v_0 r} =   \frac{1}{2} \frac{p_{\rm discard}}{p_{\rm switch}}\,.
\end{equation}
However, the discretized approximation of the telegraph process is equivalent to the continuous process only when $p_{\rm switch}\ll 1$, if $\alpha$ is not small. When $\alpha$ is small, on the other hand, then this constraint does not matter because we know that the details of the microscopic statistics are unimportant for the macroscopic statistics in this limit (see main text), provided that the effective step sizes (Kuhn lengths) of the continuous and discrete processes 
match  so that the excluded volume per step is the same. For the continuous process with $v_0=1$  the effective step size is
\begin{equation}
\int_{-\infty}^\infty \!\!{\rm d}t\,\exp(-2 r |t|)  = r^{-1}\,.
\end{equation}
For the discretized telegraph model that we simulate, the effective
step size equals 
\begin{equation} 
\label{eq:step_eff}
\ell \sum_{n=-\infty}^\infty C(n) = \ell (p_{\rm switch}^{-1} - 1)\,.
\end{equation}
Here $C(n) \equiv (1 - 2 p_{\rm switch})^{|n|}$ is the correlation function of the discretized telegraph process.
Replacing $r^{-1}$ in Eq.~(\ref{eq:ads}) by the effective
step size (\ref{eq:step_eff}), we find that in the discretized
model the scaling parameter must be computed as:
\begin{equation}
\label{eq:alpha2}
\alpha = \frac{p_{\rm discard}}{2} \frac{1 - p_{\rm switch}}{ p_{\rm switch}} \,.
\end{equation}
In the limit $p_{\rm switch}\to 0$ we obtain Eq.~(\ref{eq:ads}), as expected. In the opposite limit, $p_{\rm switch}\to 1/2$, we obtain $\alpha \to p_{\rm discard}/2$. For weak self avoidance ($p_{\rm discard}\ll 1$) this random-walk limit corresponds to small values of $\alpha$, so that the discretized
process is an accurate approximation in this limit, if we use Eq.~(\ref{eq:alpha2}).

We ensure by inspection that the large-$T$ limit was reached for the included simulations. Further, for all simulations included in the analysis, the total time $T$ obeys $ 2Tr > 30$. 

\begin{figure}[t!]
\includegraphics[clip,width=6.5in]{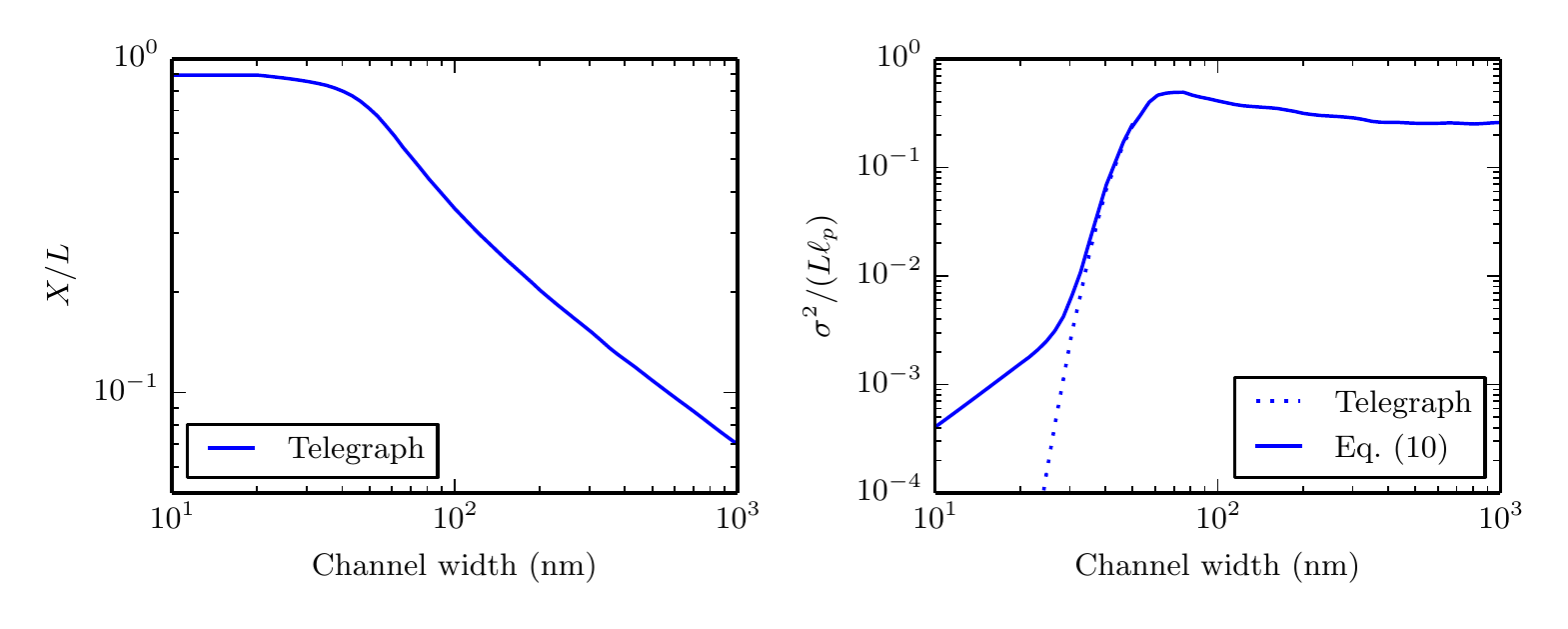}
\caption{Results of telegraph-model simulations for $\ellp$ = 50 nm and $w$ = 5 nm as a function of the channel width for the fractional extension (left) and for the normalized variance (right).
\label{fig:telegraph_expt}
}
\end{figure}

Results of our computer simulations of the telegraph model are summarized
in Fig.~4 in the main text, showing scaled extension and extension variance as a function of the scaling variable $\alpha$. These scaled plots demonstrate the universal nature of the telegraph-model predictions, as they collapse the data to a universal curve. However, 
for direct comparison of the telegraph-model predictions to experimental data on nano-confined DNA it may be practical to have the telegraph-model results
in unscaled, dimensional form. The conversion between dimensional and dimensionless variables is readily obtained from Fig.~3 in the main text. Fig.~\ref{fig:telegraph_expt} shows the result for typical values for DNA in a buffer of high ionic strength: $\ellp$ = 50 nm and $w$ = 5 nm. The results are plotted as a function of the channel width $D$.

\section{\,\,Ideal-polymer simulations}
\label{sec:ideal}
The data in Fig.~3 in the main text were obtained from simulations of confined {\em ideal} wormlike chains using 
the algorithm of Dai {\em et al}.\cite{Dai2014} 
The chain consists of $N_{\rm b}$ touching beads with diameter $d$, so
that the bond length is $\bondlength=d$. All other lengths are expressed in terms of the bond length $\bondlength$. The semiflexibility of the chain is imposed through a bending energy
\begin{equation}
\label{eq:ubend}
 U_{\mathrm{bend}} = k_B T \kappa (1 - \cos \varphi),
\end{equation}
based on the angle $\varphi$ between contiguous trios of beads. The prefactor for the bending energy, 
$\kappa$, is related to the persistence length $\ellp$ and 
bond length $\bondlength$ as \cite{Muralidhar2015}
\begin{equation}
\label{eq:kappa}
\frac{\ellp}{\bondlength} = \frac{\kappa}{\kappa - \kappa \coth(\kappa) + 1} \,,
\end{equation}
The alignment parameter $a$ and the global persistence length $g$ were calculated by fitting the simulated correlation function to Eq.~(2) in the main text. The parameter $\pen$ was computed by evaluating the average
in Eq.~(\ref{eq:penresult}) using the ideal wormlike chain simulations.  

\subsection{\,Ideal tangent correlations. Determination of $g$ and $a$ in terms of $D$ and $\ellp$}
\begin{figure}[t]
\includegraphics[clip,width=\textwidth]{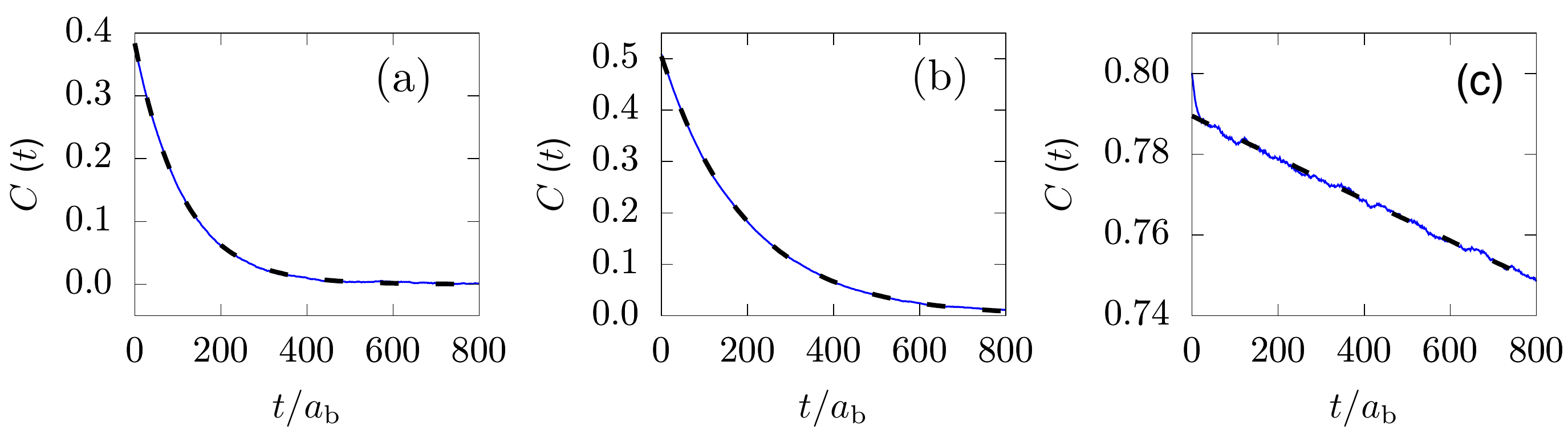}
\caption{\label{fig:correlationFunction}
Tangent correlation function measured in ideal simulations described
in Section \ref{sec:ideal} (solid blue line) and fitted to an exponential function (dashed black line). a) Wide channel ($D = 4 \ellp$). b) Intermediate channel ($D = 1.6 \ellp$). c) Narrow channel ($D = 0.43 \ellp$).  }
\end{figure}

Eq.~(2) of the main text defines the parameters $a$ and $g$ in terms of the $y$-intercept and decay length, respectively, of the exponentially decaying correlation function of the confined {\em ideal} polymer, disregarding self avoidance. We extract these parameters from ideal-polymer simulations by directly measuring the correlation function $C(t)$ of the ideal polymer, for segments far from either end of the polymer. Some example measurements are shown in Fig.~\ref{fig:correlationFunction}, together with the result of fitting the correlation function to an exponential function. Note that in the right panel for very short separations $t < \lambda$, the decay is not exponential. Otherwise the agreement is excellent.  The resulting values of $a$ and $g$ are plotted versus $D/\ellp$ in Fig.~\ref{fig:parameters_ideal} (a) and (b).
The simulations were performed by varying $D$ as well as $\ellp$. 
The parameters $a$ and $g/\ellp$ are functions of $D/\ellp$ only, since changing $D$ and $\ellp$ by the same factor has no physical significance, but is equivalent to a corresponding change in the unit of length.

\subsection{\,Determination of the parameter $\pen$ in terms of $D,w,\ellp$}
The parameter $\pen$  is obtained by evaluating in confined ideal-polymer simulations the average in Eq.~(\ref{eq:penresult}).
To this end, we discretize the delta function by dividing the channel cross-section into $N_{\rm bins}\times N_{\rm bins}$ bins of size 
${\rm d}y \times {\rm d}z$. The value of the discretized delta function is 
$({\rm d}y {\rm d}z)^{-1}$  when $\ve x_\perp$ and $\ve x_\perp'$ 
are in the same bin, and zero otherwise. To approximate the average 
in Eq.~(\ref{eq:penresult}), we increment a counter $c$ by 
$\sin(\theta) /({\rm d}y {\rm d}z)$ whenever two segments are in the same bin. 
Our estimate of $\pen$ is then given by $c \times 2w / N_{\rm iter}$, 
where $N_{\rm iter}$ is the number of iterations in the simulations.

Note that a correct computation of the average  requires that the two segments 
are well separated from both each other, and from either end of the polymer. Here, well separated means that the contour separation should be large enough that the two positions are statistically independent. This is the case if the contour length separation  is
much larger than $D^{2/3} \ellp^{1/3} + D^2 / \ellp$.\cite{Chen2013} To ensure that this is indeed the case in simulations, the contour length between 
the two segments was kept at either 30 or 60 times this limit. To avoid end effects, we further required that both segments in the pair  were located at least a contour separation of $L/3$ from the left and right end, respectively.

Further, the number of bins in the discretization of the average must be large enough that the discretization error is negligible. In simulations we divided the channel cross section into $40 \times 40$, $60 \times 60$, or $80 \times 80$ bins. No significant difference was observed.

Finally, to avoid discretization error in the polymer configuration, the bead 
diameter $d$ must be significantly smaller than the channel width $D$. In simulations we ensured that $D > 40 d$. Again, we observed no effect from changing $D / d$.
\begin{figure}[t]
\begin{overpic}[width=1.0\textwidth]{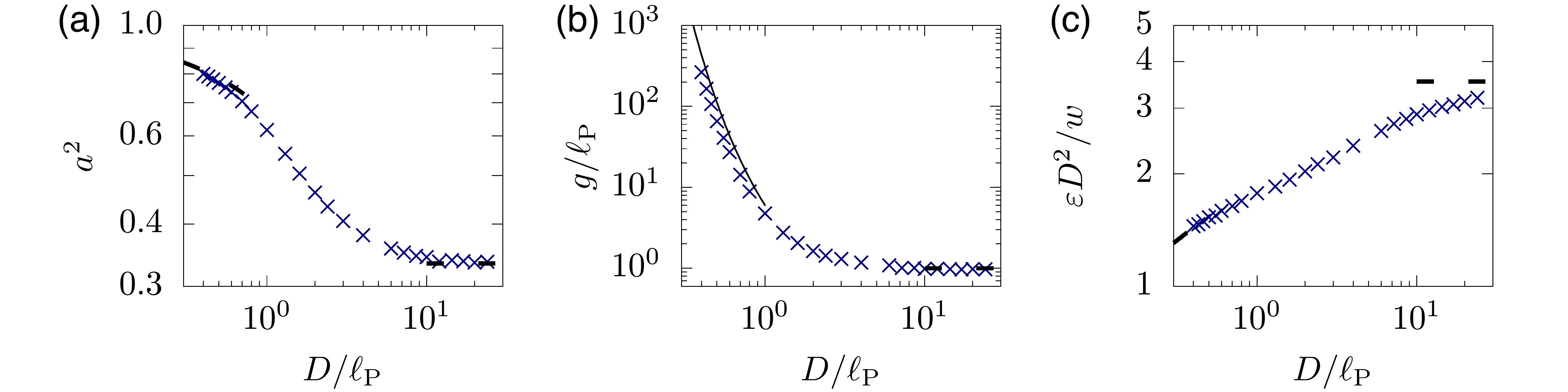}
\put(67.2,11.5){\rotatebox{90}{\colorbox{white}{$\varepsilon D^2/w$}}}
\end{overpic}
\caption{\label{fig:parameters_ideal}
Dependence of $a$, $g$, and $\pen$ upon $D/\ellp$. 
Results of ideal-polymer simulation as described in the text.
The solid line is the result from Muralidhar {\it et al.}\cite{Muralidhar2014}
The dashed lines represent different asymptotic limits discussed in this Supplemental Material.
Fitting the constant $C$ in Eq.~(\ref{eq:lexell})
to the ideal-polymer simulations gives $C\approx 1.95$. }
\end{figure}
Fig.~\ref{fig:parameters_ideal}(c) shows the result for $\pen$ as a function of $D/\ellp$, 
computed from these ideal-polymer simulations (symbols).
To obtain Fig.~\ref{fig:parameters_ideal}(c), $D$ and $\ellp$ were varied. 
A similar argument as in the previous Section shows that $\pen D^2/w$ is a function of $D/\ellp$ only.

\subsection{\,Known asymptotes of $g, a$ and $p$
for small and large $D/\ellp$}
Now we discuss the asymptotes shown in Fig.~\ref{fig:parameters_ideal}.
In panel (a) the dashed line for small values of $D/\ellp$ is the Odijk result
for $X^2/L^2$ as computed by Burkhardt {\em et al.}\cite{burkhardt2010}. In panel (c) the dashed line for small values of $D/\ellp$
is the prediction in the backfolded Odijk regime.\cite{Odijk2008,Muralidhar2014} The dashed lines for large values of $D/\ellp$ are the predictions for the extended de Gennes regime,\cite{Werner2014} as given in Section \ref{sec:limits}. For the global persistence length, we also compared our results to the previous simulations by Muralidhar {\it et al.,}\cite{Muralidhar2014} which were obtained using a different method, and a slightly different definition of $g$, 
solid line in Fig.~\ref{fig:parameters_ideal}(b). We find good agreement.

\section{\,\,Direct numerical simulations of the three-dimensional self\--avoi\-ding polymer in confinement}
\label{sec:PERM}
We have thus far explained how we simulated the telegraph model (in \S \ref{sec:simulation_telegraph}) and how we computed the parameter $\alpha$ from ideal chain simulations (in \S \ref{sec:ideal}). We now proceed to describe the most computationally expensive simulations, namely the direct numerical simulation of a wormlike chain confined in a nanochannel with excluded volume. 
The direct numerical simulation (DNS) data in Fig.~4 in the main text were obtained by simulating a confined {\em self-avoiding} discrete wormlike chain via the Pruned Enriched Rosenbluth Method (PERM) \cite{Grassberger1997,Prellberg2004}.  
The polymer model is similar to the one used for the simulations of ideal polymers
in Section \ref{sec:ideal}, with the important difference that we now take self avoidance into account. We have previously developed highly efficient simulation methods for this problem \cite{Tree2013a,Tree2013} that we modified here to more accurately reflect the cylindrical excluded volume. 

The bending energy is given by Eq.~(\ref{eq:ubend}), and the persistence
length is determined by Eq.~(\ref{eq:kappa}).
As in Section \ref{sec:ideal}, the chain consists of $N_{\rm b}$ 
touching beads of diameter $d$.
Bead-bead excluded volume is treated as an infinite penalty if the center-to-center distance between the beads is less 
their diameter (representing the effective width $w$ of the polymer).  To approximate the cylindrical excluded volume interactions while retaining good resolution of the bending of the chain, we use $d/\bondlength = 2$, rather than the customary $d/\bondlength = 1$. The overlap penalty was not included for contiguous beads along the backbone that would overlap due to the choice $d > \bondlength$. A hard-core excluded volume penalty is imposed also for bead-wall overlap, such that the beads are confined inside a channel of size $D_{\mathrm{eff}} = D - w$.
For a given parameter set $(\ellp,D)$, we grew chains using an off-lattice PERM algorithm for confined polymers\cite{Tree2013a,MuraidharThesis} to $N_{\rm b}$ = 40,000 beads, using 10 independent simulations with 100,000 tours per simulation. 
The values of $X$ and $\sigma^2$ represent the average over these 10 sets, and we use the standard error of the mean computed from these independent sets to estimate the sampling error. In many cases, the error is smaller than the symbol size.  For each pair ($\ellp,D$), we confirmed that both $X$ and $\sigma^2$ achieved linear scaling with $L$.
Table \ref{table:PERM} lists all combinations of the persistence length and channel width used to produce the data in Fig.~4 of the main text.

\setlength{\tabcolsep}{20pt}
\renewcommand{\arraystretch}{0.6}

\begin{longtable}{ c c  c  c  c c c }
\caption{List of parameters for PERM simulations. 
\label{table:PERM}
}\\

\hline \hline
\\
\multicolumn{1}{c}{{$\ellp$}} & \multicolumn{1}{c}{{$w$}} &  \multicolumn{1}{c}{{$D$}}  \\
\\
 \hline \\
\endfirsthead

\multicolumn{3}{c}%
{{ \tablename\ \thetable{} -- continued from previous page}} \\
\multicolumn{1}{c}{{$\ellp$}} & \multicolumn{1}{c}{{$w$}} &  \multicolumn{1}{c}{{$D$ }} \\ \\ \hline \\
\endhead

\\ \hline  \multicolumn{3}{r}{{Continued on next page}} \\ 
\endfoot

\hline \hline
\endlastfoot
24.00 & 2 & 11.6 \\
24.00 & 2 & 12.32 \\
24.00 & 2 & 13.04 \\
24.00 & 2 & 14 \\
24.00 & 2 & 15.2 \\
24.00 & 2 & 16.4 \\
24.00 & 2 & 18.8 \\
24.00 & 2 & 21.2 \\
24.00 & 2 & 26 \\
24.00 & 2 & 32 \\
24.00 & 2 & 40.4 \\
24.00 & 2 & 50 \\
24.00 & 2 & 59.6 \\
24.00 & 2 & 74 \\
24.00 & 2 & 98 \\
24.00 & 2 & 146 \\
48.00 & 2 & 21.2 \\
48.00 & 2 & 22.64 \\
48.00 & 2 & 24.08 \\
48.00 & 2 & 26 \\
48.00 & 2 & 28.4 \\
48.00 & 2 & 30.8 \\
48.00 & 2 & 35.6 \\
48.00 & 2 & 40.4 \\
48.00 & 2 & 50 \\
48.00 & 2 & 62 \\
48.00 & 2 & 78.8 \\
48.00 & 2 & 98 \\
48.00 & 2 & 117.2 \\
48.00 & 2 & 146 \\
48.00 & 2 & 194 \\
48.00 & 2 & 290 \\
72.00 & 2 & 30.8 \\
72.00 & 2 & 32.96 \\
72.00 & 2 & 35.12 \\
72.00 & 2 & 38 \\
72.00 & 2 & 41.6 \\
72.00 & 2 & 45.2 \\
72.00 & 2 & 52.4 \\
72.00 & 2 & 59.6 \\
72.00 & 2 & 74 \\
72.00 & 2 & 92 \\
72.00 & 2 & 117.2 \\
72.00 & 2 & 146 \\
72.00 & 2 & 174.8 \\
72.00 & 2 & 218 \\
72.00 & 2 & 290 \\
72.00 & 2 & 434 \\
18.95 & 2 & 18.23 \\
9.34 & 2 & 14.99 \\
12.27 & 2 & 21.46 \\
18.90 & 2 & 36.36 \\
24.55 & 2 & 49.24 \\
6.17 & 2 & 16.34 \\
9.35 & 2 & 30.02 \\
12.32 & 2 & 43.14 \\
18.97 & 2 & 73.02 \\
24.50 & 2 & 98.23 \\
7.78 & 2 & 23.24 \\
9.45 & 2 & 30.45 \\
12.91 & 2 & 45.76 \\
16.79 & 2 & 63.18 \\
15.02 & 2 & 68.97 \\
4.95 & 2 & 14.09 \\
9.79 & 2 & 12.78 \\
9.79 & 2 & 21.55 \\
9.79 & 2 & 57.99 \\
9.79 & 2 & 70.28 \\
11.13 & 2 & 33.47 \\
11.13 & 2 & 33.67 \\
11.13 & 2 & 33.95 \\
11.13 & 2 & 34.47 \\
6.03 & 2 & 12.03 \\
6.03 & 2 & 12.15 \\
11.13 & 2 & 28.78 \\
11.13 & 2 & 37.36 \\
13.60 & 2 & 46.29 \\
13.60 & 2 & 52.91 \\
21.75 & 2 & 43.88 \\
21.75 & 2 & 104.85 \\
21.75 & 2 & 128.47 \\
7.35 & 2 & 34.65 \\
7.35 & 2 & 37.03 \\
7.35 & 2 & 39.25 \\
5.95 & 2 & 23.21 \\
11.26 & 2 & 57.88 \\
16.99 & 2 & 96.48 \\
22.37 & 2 & 133.12 \\
16.99 & 2 & 43.08 \\
18.21 & 2 & 42.59 \\
11.26 & 2 & 23.50 \\
7.21 & 2 & 12.68 \\

\hline 
\end{longtable}%

\eject

\section{\,\,Summary of experimental data analyzed in main text}
Here we explain how the experimental data sets in Figs.~1 and 4 in the main
text  were selected, and how the effective channel width, persistence length $\ellp$, and effective $w$ were computed from the experimental parameters.

\subsection{\,Computation of contour length, persistence length and effective polymer width}

All the selected experiments presented in Figs.~1 and 4 in the main text used bis-intercalating dyes (TOTO-1 or YOYO-1) for visualization of DNA. These dyes are expected to increase the contour length, $L$ linearly in proportion to the amount of dye bound. In order to calculate $L$, we assume that every bound dye molecule adds 0.44 nm to the bare contour length of 0.34 nm per base pair.\cite{Gupta2014} Note that the intercalation effect was neglected for Reinhart {\it et al.},\cite{Reinhart2015} who  used a very low dye ratio (1 dye molecule per 40 base pairs).

There has been a substantial debate about the effect of bis-intercalating dyes on persistence length of DNA, $\ellp$, as summarized by Kundukad {\it et al.}\cite{Kundukad2014} We concur with their conclusion that the persistence length is at best weakly affected by YOYO if the system is equilibrated. Therefore, we only account for the effect of ionic strength of DNA solutions when computing the persistence length. The ionic strength for the various buffers used in these experiments were calculated using a MATLAB program developed in our previous work.\cite{Gupta2014} Thereafter, $\ellp$ was calculated using Dobrynin's empirical formula\cite{Dobrynin2006} which has been experimentally validated. \cite{Hsieh2008} The effective polymer width $w$ was computed from Stigter's theory.\cite{Stigter1977}

\subsection{\,Tabulation of experimental data}
Fig.~1 in the main text reports the parameters from experiments for cases where the aspect ratio is no greater than 1.5 and $\ellp$/$w$ is no less than 2 to ensure (i) that the channels are close to square, in order to correspond to our simulations, and (ii) that the DNA can be modeled as  semiflexible chains. The raw data appearing in that Figure are tabulated in Table \ref{table:Fig2}. The effective channel width is computed by 
\begin{equation}
D_{\mathrm{eff}} = \sqrt{(D_1-w)(D_2-w)}
\end{equation}
where $D_1$ and $D_2$ are the two dimensions of the channel and $w$ is an approximation for the excluded volume due to DNA-wall 
electrostatic interactions.

\setlength{\tabcolsep}{8pt}
\renewcommand{\arraystretch}{0.6}

\begin{longtable}{ c c  c  c  c c c }
\caption{Experiments in Fig.~1 of the main text.  
\label{table:Fig2}
}\\

\hline \hline
\\
\multicolumn{1}{c}{{Source}} & \multicolumn{1}{c}{{Ionic Strength (mM)}} &  \multicolumn{1}{c}{{$\ellp$ (nm)}} &  \multicolumn{1}{c}{{$w$ (nm)}} &  \multicolumn{1}{c}{{$D_1$ (nm) }} & \multicolumn{1}{c}{{$D_2$ (nm) }} \\
\\
 \hline \\
\endfirsthead

\multicolumn{7}{c}%
{{ \tablename\ \thetable{} -- continued from previous page}} \\
\multicolumn{1}{c}{{Source}} & \multicolumn{1}{c}{{Ionic Strength (mM)}} &  \multicolumn{1}{c}{{$\ellp$ (nm)}} &  \multicolumn{1}{c}{{$w$ (nm)}}  &  \multicolumn{1}{c}{{$D_1$ (nm) }} & \multicolumn{1}{c}{{$D_2$ (nm) }} \\ \\ \hline \\
\endhead

\\ \hline  \multicolumn{7}{r}{{Continued on next page}} \\ 
\endfoot

\hline \hline
\endlastfoot

\\
 \\
 Reisner {\it et al.}\cite{Reisner2005} & 16 & 61 & 13 & 30 & 40 \\
 & 16 & 61 & 13 & 60 & 80 \\
 & 16 & 61 & 13 & 80 & 80 \\
 & 16 & 61 & 13 & 130 & 140 \\
 & 16 & 61 & 13 & 300 & 440 \\
 & 16 & 61 & 13 & 440 & 440 \\
 \\
Reisner {\it et al.}\cite{Reisner2007} & 4 & 76 & 25 & 50 & 50 \\
 & 14 & 62 & 13 & 50 & 50 \\
 & 30 & 57 & 9 & 50 & 50 \\
 & 107 & 52 & 5 & 50 & 50 \\
 & 4 & 76 & 25 & 100 & 100 \\
 & 14 & 62 & 13 & 100 & 100 \\
 & 30 & 57 & 9 & 100 & 100 \\
 & 106 & 52 & 6 & 100 & 100 \\
 & 261 & 50 & 4 & 100 & 100 \\
 & 4 & 75 & 24 & 200 & 200 \\
 & 14 & 62 & 13 & 200 & 200 \\
 & 30 & 57 & 9 & 200 & 200 \\
 & 107 & 52 & 5 & 200 & 200 \\
 & 259 & 50 & 4 & 200 & 200 \\
 & 8 & 67 & 17 & 200 & 200 \\
 & 14 & 62 & 13 & 200 & 200 \\
 & 35 & 56 & 9 & 200 & 200 \\
 & 74 & 53 & 6 & 200 & 200 \\
 \\
Thamdrup {\it et al.}\cite{Thamdrup2008} & 53 & 54 & 7 & 250 & 250 \\
 \\
Zhang {\it et al.}\cite{Zhang2008} & 1 & 106 & 53 & 200 & 300 \\
 & 2 & 89 & 36 & 200 & 300 \\
 & 3 & 83 & 31 & 200 & 300 \\
 & 10 & 65 & 15 & 200 & 300 \\
 & 27 & 58 & 10 & 200 & 300 \\
 & 33 & 57 & 9 & 200 & 300 \\
 & 1 & 106 & 53 & 300 & 300 \\
 & 2 & 89 & 36 & 300 & 300 \\
 & 3 & 83 & 31 & 300 & 300 \\
 & 11 & 65 & 15 & 300 & 300 \\
 & 27 & 58 & 10 & 300 & 300 \\
 & 33 & 57 & 9 & 300 & 300 \\
 \\
Utko {\it et al.}\cite{Utko2011} & 23 & 59 & 11 & 150 & 208 \\
 & 23 & 59 & 11 & 150 & 221 \\
 \\
Kim {\it et al.}\cite{Kim2011} & 1 & 103 & 50 & 250 & 250 \\
 & 2 & 88 & 35 & 250 & 250 \\
 \\
Werner {\it et al.}\cite{Werner2012} & 4 & 76 & 25 & 180 & 121 \\
 & 4 & 76 & 25 & 180 & 259 \\
 & 23 & 59 & 11 & 180 & 128 \\
 & 23 & 59 & 11 & 180 & 262 \\
 & 40 & 56 & 8 & 180 & 123 \\
 & 40 & 56 & 8 & 180 & 261 \\
 \\

Gupta {\it et al.}\cite{Gupta2014} & 7 & 69 & 19 & 100 & 100 \\
 \\
Reinhart {\it et al.}\cite{Reinhart2015} & 103 & 52 & 6 & 40 & 40 \\
 & 103 & 52 & 6 & 42 & 42 \\
 & 103 & 52 & 6 & 43 & 43 \\
 & 103 & 52 & 6 & 49 & 49 \\
 & 103 & 52 & 6 & 51 & 51 \\
 \\
Gupta {\it et al.}\cite{Gupta2015} & 7 & 69 & 19 & 300 & 350 \\
 & 7 & 69 & 19 & 300 & 450 \\
 \\
Iarko {\it et al.}\cite{Iarko2015} & 4 & 77 & 26 & 300 & 302 \\
 & 24 & 59 & 10 & 300 & 302 \\
 & 77 & 53 & 6 & 300 & 302 \\
 & 184 & 51 & 5 & 300 & 302 \\
 & 4 & 77 & 26 & 120 & 150 \\
 & 77 & 53 & 6 & 120 & 151 \\
 \\
Alizadehheidari {\it et al.}\cite{Alizadehheidari2015} & 95 & 52 & 6 & 100 & 150 \\
 & 24 & 59 & 10 & 100 & 150 \\
 & 7 & 69 & 19 & 100 & 150 \\
 & 4 & 77 & 26 & 100 & 100 \\

\\
\hline 
\end{longtable}%

\eject
Fig.~4 in the main text reports values of $X$ and $\sigma^2$ from experiments for cases where the aspect ratio is no greater than 1.5 and $\ellp$/$w$ is no less than 2. Table \ref{table:Fig4} provides a tabulated list of the experimental parameters and data.

\setlength{\tabcolsep}{8pt}
\renewcommand{\arraystretch}{0.6}
\begin{longtable}{ c c  c  c  c c c }
\caption{Experimental data in Fig.~4 of the main text.
The values of $\alpha$ marked with an asterisk are obtained
by extrapolation.
\label{table:Fig4}
}\\

\hline \hline
\\
\multicolumn{1}{c}{{Source}} & \multicolumn{1}{c}{{$\ellp$/$w$}} &  \multicolumn{1}{c}{{$D_{\mathrm{eff}}$/$\ellp$}} & \multicolumn{1}{c}{{$\alpha$}} & \multicolumn{1}{c}{{ $X$/$L$}} & \multicolumn{1}{c}{{$\sigma^2$/($L$$\ellp$)}} & \multicolumn{1}{c}{{$L$ ($\mu$m)}} \\
\\
 \hline \\
\endfirsthead

\multicolumn{7}{c}%
{{ \tablename\ \thetable{} -- continued from previous page}} \\
\hline \\ \multicolumn{1}{c}{{Source}} & \multicolumn{1}{c}{{$\ellp$/$w$}} & \multicolumn{1}{c}{{$D_{\mathrm{eff}}$/$\ellp$}} & \multicolumn{1}{c}{{$\alpha$}} & \multicolumn{1}{c}{{ $X$/$L$}} & \multicolumn{1}{c}{{$\sigma^2$/($L$$\ellp$)}}  & \multicolumn{1}{c}{{$L$ ($\mu$m)}} \\ \\ \hline \\
\endhead

\\ \hline  \multicolumn{7}{r}{{Continued on next page}} \\ 
\endfoot

\hline \hline
\endlastfoot

Reisner {\it et al.}\cite{Reisner2005} & 4.9 & 0.4 & 1543$^\ast$ & 0.73 &  & 18.6 \\
 & 4.9 & 0.9 & 3.05 & 0.63 &  & 18.6 \\
 & 4.9 & 1.1 & 1.50 & 0.50 &  & 18.6 \\
 & 4.9 & 2.0 & 0.25 & 0.36 &  & 18.6 \\
 & 4.9 & 5.7 & 0.03 & 0.15 &  & 18.6 \\
 & 4.9 & 7.0 & 0.02 & 0.12 &  & 18.6 \\
 \\
Reisner {\it et al.}\cite{Reisner2007} & 3.1 & 0.3 & 5119$^\ast$ & 0.82 &  & 21.0 \\
 & 4.7 & 0.6 & 32.9 & 0.82 &  & 21.0 \\
 & 6.1 & 0.7 & 8.44 & 0.74 &  & 21.0 \\
 & 9.5 & 0.9 & 2.20 & 0.60 &  & 21.0 \\
 & 3.1 & 1.0 & 3.59 & 0.73 &  & 21.0 \\
 & 4.7 & 1.4 & 0.70 & 0.55 &  & 21.0 \\
 & 6.1 & 1.6 & 0.36 & 0.45 &  & 21.0 \\
 & 9.5 & 1.8 & 0.16 & 0.60 &  & 21.0 \\
 & 12.3 & 1.9 & 0.11 & 0.32 &  & 21.0 \\
 & 3.1 & 2.3 & 0.28 & 0.45 &  & 21.0 \\
 & 4.7 & 3.0 & 0.11 & 0.26 &  & 21.0 \\
 & 6.2 & 3.3 & 0.07 & 0.23 &  & 21.0 \\
 & 9.5 & 3.7 & 0.03 & 0.17 &  & 21.0 \\
 & 12.2 & 3.9 & 0.02 & 0.15 &  & 21.0 \\
 & 3.9 & 2.7 & 0.16 & 0.26 &  & 21.0 \\
 & 4.7 & 3.0 & 0.11 & 0.27 &  & 21.0 \\
 & 6.5 & 3.4 & 0.06 & 0.20 &  & 21.0 \\
 & 8.4 & 3.6 & 0.04 & 0.16 &  & 21.0 \\
 \\
Thamdrup {\it et al.}\cite{Thamdrup2008} & 7.5 & 4.5 & 0.03 & 0.19 & 0.26 & 70.9 \\
 \\
Zhang {\it et al.}\cite{Zhang2008} & 2.0 & 1.8 & 0.79 & 0.23 &  & 59.5 \\
 & 2.4 & 2.3 & 0.34 & 0.21 &  & 59.5 \\
 & 2.6 & 2.6 & 0.27 & 0.19 &  & 59.5 \\
 & 4.2 & 3.5 & 0.09 & 0.19 &  & 59.5 \\
 & 5.9 & 4.1 & 0.05 & 0.17 &  & 59.5 \\
 & 6.3 & 4.2 & 0.04 & 0.19 &  & 59.5 \\
 & 2.0 & 2.3 & 0.43 & 0.19 &  & 59.5 \\
 & 2.4 & 3.0 & 0.21 & 0.17 &  & 59.5 \\
 & 2.7 & 3.2 & 0.16 & 0.17 &  & 59.5 \\
 & 4.2 & 4.4 & 0.06 & 0.16 &  & 59.5 \\
 & 5.9 & 5.0 & 0.03 & 0.14 &  & 59.5 \\
 & 6.3 & 5.1 & 0.03 & 0.15 &  & 59.5 \\
 \\
Utko {\it et al.}\cite{Utko2011} & 5.6 & 2.8 & 0.10 & 0.31 &  & 18.6 \\
 & 5.6 & 2.8 & 0.10 & 0.29 &  & 18.6 \\
 & 5.6 & 2.9 & 0.10 & 0.31 &  & 18.6 \\
 & 5.6 & 2.9 & 0.10 & 0.31 &  & 18.6 \\
 \\
Kim {\it et al.}\cite{Kim2011} & 2.1 & 1.9 & 0.64 & 0.62 &  & 21.8 \\
 & 2.5 & 2.4 & 0.31 & 0.57 &  & 21.8 \\
 \\
Werner {\it et al.}\cite{Werner2012} & 3.0 & 1.6 & 0.73 & 0.53 &  & 18.6 \\
 & 3.0 & 1.6 & 0.71 & 0.47 &  & 18.6 \\
 & 3.0 & 1.7 & 0.65 & 0.48 &  & 18.6 \\
 & 3.0 & 1.6 & 0.67 & 0.50 &  & 18.6 \\
 & 3.0 & 1.7 & 0.63 & 0.51 &  & 18.6 \\
 & 3.0 & 1.7 & 0.59 & 0.48 &  & 18.6 \\
 & 3.0 & 1.8 & 0.56 & 0.47 &  & 18.6 \\
 & 3.0 & 1.8 & 0.53 & 0.45 &  & 18.6 \\
 & 3.0 & 1.8 & 0.50 & 0.46 &  & 18.6 \\
 & 3.0 & 1.8 & 0.51 & 0.44 &  & 18.6 \\
 & 3.0 & 1.9 & 0.47 & 0.44 &  & 18.6 \\
 & 3.0 & 1.9 & 0.47 & 0.46 &  & 18.6 \\
 & 3.0 & 1.9 & 0.45 & 0.44 &  & 18.6 \\
 & 3.0 & 2.0 & 0.42 & 0.42 &  & 18.6 \\
 & 3.0 & 2.0 & 0.39 & 0.41 &  & 18.6 \\
 & 3.0 & 2.1 & 0.38 & 0.39 &  & 18.6 \\
 & 3.0 & 2.1 & 0.37 & 0.41 &  & 18.6 \\
 & 3.0 & 2.1 & 0.35 & 0.41 &  & 18.6 \\
 & 3.0 & 2.2 & 0.34 & 0.39 &  & 18.6 \\
 & 3.0 & 2.2 & 0.32 & 0.40 &  & 18.6 \\
 & 3.0 & 2.2 & 0.31 & 0.38 &  & 18.6 \\
 & 3.0 & 2.3 & 0.30 & 0.39 &  & 18.6 \\
 & 3.0 & 2.3 & 0.29 & 0.37 &  & 18.6 \\
 & 3.0 & 2.4 & 0.27 & 0.38 &  & 18.6 \\
 & 3.0 & 2.4 & 0.27 & 0.37 &  & 18.6 \\
 & 3.0 & 2.4 & 0.26 & 0.35 &  & 18.6 \\
 & 3.0 & 2.4 & 0.25 & 0.37 &  & 18.6 \\
 & 3.0 & 2.5 & 0.25 & 0.34 &  & 18.6 \\
 & 3.0 & 2.4 & 0.26 & 0.34 &  & 18.6 \\
 & 5.6 & 2.4 & 0.14 & 0.31 &  & 18.6 \\
 & 5.6 & 2.5 & 0.14 & 0.30 &  & 18.6 \\
 & 5.6 & 2.5 & 0.13 & 0.32 &  & 18.6 \\
 & 5.6 & 2.5 & 0.13 & 0.31 &  & 18.6 \\
 & 5.6 & 2.6 & 0.12 & 0.31 &  & 18.6 \\
 & 5.6 & 2.7 & 0.11 & 0.29 &  & 18.6 \\
 & 5.6 & 2.8 & 0.10 & 0.28 &  & 18.6 \\
 & 5.6 & 2.9 & 0.10 & 0.28 &  & 18.6 \\
 & 5.6 & 3.0 & 0.09 & 0.28 &  & 18.6 \\
 & 5.6 & 3.1 & 0.09 & 0.26 &  & 18.6 \\
 & 5.6 & 3.1 & 0.09 & 0.25 &  & 18.6 \\
 & 5.6 & 3.2 & 0.08 & 0.28 &  & 18.6 \\
 & 5.6 & 3.2 & 0.08 & 0.27 &  & 18.6 \\
 & 5.6 & 3.3 & 0.07 & 0.26 &  & 18.6 \\
 & 5.6 & 3.4 & 0.07 & 0.24 &  & 18.6 \\
 & 5.6 & 3.5 & 0.07 & 0.25 &  & 18.6 \\
 & 5.6 & 3.5 & 0.07 & 0.24 &  & 18.6 \\
 & 6.8 & 2.5 & 0.11 & 0.29 &  & 18.6 \\
 & 6.8 & 2.6 & 0.10 & 0.28 &  & 18.6 \\
 & 6.8 & 2.6 & 0.10 & 0.26 &  & 18.6 \\
 & 6.8 & 2.8 & 0.09 & 0.27 &  & 18.6 \\
 & 6.8 & 2.8 & 0.08 & 0.26 &  & 18.6 \\
 & 6.8 & 3.0 & 0.08 & 0.26 &  & 18.6 \\
 & 6.8 & 3.1 & 0.07 & 0.25 &  & 18.6 \\
 & 6.8 & 3.1 & 0.07 & 0.23 &  & 18.6 \\
 & 6.8 & 3.3 & 0.06 & 0.24 &  & 18.6 \\
 & 6.8 & 3.3 & 0.06 & 0.24 &  & 18.6 \\
 & 6.8 & 3.4 & 0.06 & 0.23 &  & 18.6 \\
 & 6.8 & 3.8 & 0.05 & 0.21 &  & 18.6 \\
 & 6.8 & 3.6 & 0.05 & 0.22 &  & 18.6 \\
 & 6.8 & 3.7 & 0.05 & 0.21 &  & 18.6 \\
 \\

Gupta {\it et al.}\cite{Gupta2014} & 3.7 & 1.2 & 1.53 & 0.81 & 0.21 & 18.6 \\
 \\
Reinhart {\it et al.}\cite{Reinhart2015} & 9.3 & 0.66 & 8.40 & 0.89 & 0.03 \\
 & 9.3 & 0.70 & 6.15 & 0.88 & 0.02 \\
 & 9.3 & 0.72 & 5.32 & 0.87 & 0.04 \\
 & 9.3 & 0.83 & 2.53 & 0.84 & 0.09 \\
 & 9.3 & 0.87 & 2.06 & 0.81 & 0.13 \\
 \\
Gupta {\it et al.}\cite{Gupta2015} & 3.7 & 4.4 & 0.06 & 0.28 & 0.26 & 63.6 \\
 & 3.7 & 4.8 & 0.06 & 0.26 & 0.27 & 63.6 \\
 & 3.7 & 5.1 & 0.05 & 0.25 & 0.27 & 63.6 \\
 \\
Iarko {\it et al.}\cite{Iarko2015} & 3.0 & 3.6 & 0.12 & 0.27 & 0.24 & 63.6 \\
 & 5.6 & 5.0 & 0.03 & 0.22 & 0.32 & 62.4 \\
 & 8.5 & 5.6 & 0.02 & 0.18 & 0.31 & 60.9 \\
 & 11.2 & 5.9 & 0.01 & 0.16 & 0.33 & 58.9 \\
 & 3.0 & 1.4 & 1.08 & 0.58 & 0.25 & 63.6 \\
 & 8.5 & 2.4 & 0.09 & 0.31 & 0.45 & 57.9 \\
 \\
Alizadehheidari {\it et al.}\cite{Alizadehheidari2015} & 9.1 & 2.2 & 0.10 &  & 0.29 & 16.2 \\
 & 5.6 & 1.9 & 0.24 &  & 0.23 & 16.2 \\
 & 3.6 & 1.5 & 0.75 &  & 0.19 & 16.2 \\
 & 3.0 & 1.0 & 4.24 &  & 0.12 & 16.2 \\

\\

\hline 
\end{longtable}%

\subsection{\,Experimental parameters in relation to asymptotic scaling regimes}
\label{sec:boundaries}

\begin{figure}[t!]
\begin{center}
\includegraphics[clip,width=5in]{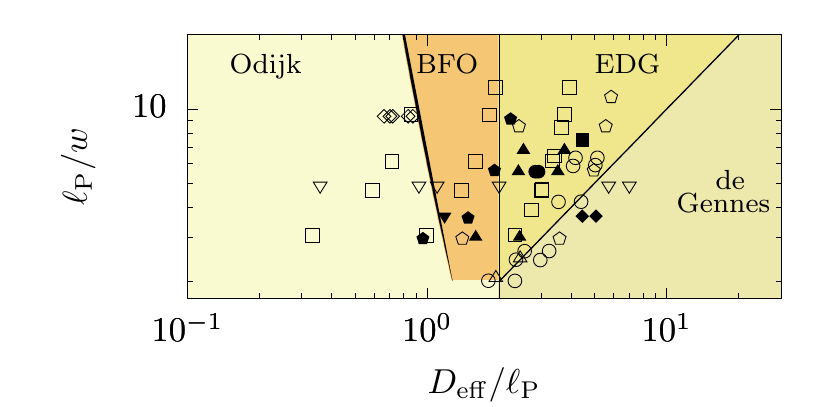}
\end{center}
\caption{Same data as Fig.~1 of the main text, here with the boundaries between the different  asymptotic regimes that are mentioned
in Section \ref{sec:boundaries}, solid lines.  \label{fig:regimes} }
\end{figure}

In the main text, we claim that the disagreement noted earlier between experiments and 
the scalings $X\sim D^{-2/3}$ or $X \sim \xi^{1/3}$ reflects that the DNA experiments do not satisfy the strong inequalities $D\gg\ellp$ or $D\ll \ellp$. To support this claim, Fig.~\ref{fig:regimes} reproduces the data in Fig.~1 of the main text but also indicates different asymptotic scaling regimes:\cite{Odijk2008} the Odijk regime; 
backfolded Odijk (BFO) regime; extended de Gennes (EDG) regime; and de Gennes regime. 
The condition for the boundary between the Odijk regime and the BFO regime is\cite{Odijk2008,Muralidhar2014} $\xi = 1$, corresponding to Eq.~(\ref{eq:xi_Odijk}) with $C$ = 1. The boundary between the BFO and 
EDG regimes is\cite{Odijk2008} $D_{\mathrm{eff}} = 2 \ellp$. The boundary between the EDG and the de Gennes regimes is $D_{\mathrm{eff}} = \ellp^2/w$, as stated in the main text. 
Fig.~\ref{fig:regimes} demonstrates that most recent experiments on nanoconfined DNA are performed close to these boundaries separating the different asymptotic regimes. 
Near the boundaries, that is for most experiments shown in Fig.~\ref{fig:regimes}, existing scaling laws must fail to describe the extension of DNA in a nanochannel.